\documentclass{nature}
\usepackage{graphicx}
\usepackage{mathabx}
\makeatletter
\renewcommand\footnotesize{%
   \@setfontsize\footnotesize\@ixpt{8}%
   \abovedisplayskip 8\p@ \@plus2\p@ \@minus4\p@
   \abovedisplayshortskip \z@ \@plus\p@
   \belowdisplayshortskip 4\p@ \@plus2\p@ \@minus2\p@
   \def\@listi{\leftmargin\leftmargini
               \topsep 4\p@ \@plus2\p@ \@minus2\p@
               \parsep 2\p@ \@plus\p@ \@minus\p@
               \itemsep \parsep}%
   \belowdisplayskip \abovedisplayskip
}
\makeatother

\title{Accreting Protoplanets in the LkCa 15 Transition Disc}

\author{Sallum, S.$^{1}$, Follette, K. B.$^{1,2}$, Eisner, J. A.$^1$, Close, L. M.$^1$, Hinz, P.$^1$, Kratter, K.$^{1}$, Males, J.$^{1}$, Skemer, A.$^1$, Macintosh, B.$^{2}$, Tuthill, P.$^3$, Bailey, V.$^1$, Defr\`{e}re, D.$^1$, Morzinski, K.$^{1}$, Rodigas, T.$^4$, Spalding, E.$^1$, Vaz, A.$^1$, Weinberger, A. J.$^{4}$ \footnote{Author Contributions: This work merged two independently acquired and analysed data sets. S.S. led preparation of the manuscript, the orbital fits, and the acquisition and analysis of the LBT data while K.B.F. led the acquisition
and analysis of the MagAO data, development of the MagAO SDI pipeline, and drafted MagAO manuscript sections. S.S., K.B.F., J.E., L.C., P.H., A.S., J.M., and K.M. contributed to one or both observing proposals. J.E. modelled circumplanetary disk and hot-start scenarios, developed the NRM mode at LBT, and supervised effort of S.S.; L.C. carried out H$\alpha$ luminosity calculations and oversaw the MagAO effort. P.H. led LBTI development and support, and helped commission the NRM mode at LBT. K.K. carried out orbital stability analysis. J.M. developed the KLIP code used in MagAO data analysis. P.T. helped develop the NRM mode at LBT. B.M. supervised the effort of K.B.F.; S.S., K.B.F., J.E., L.C., and K.K. contributed
key aspects of the manuscript. A.S., V.B., D.D., E.S., and A.V. supported the LBT observations. J.M., K.M., T.R., and A.W. supported the MagAO observations.}}

\begin{document}

\maketitle

\begin{affiliations}
 \item Astronomy Department, University of Arizona, 933 North Cherry Avenue, Tucson, AZ 85721, USA
 \item Kavli Institute for Particle Astrophysics and Cosmology, Stanford University, Stanford, CA 94305
 \item School of Physics, University of Sydney, Sydney, NSW 2006, Australia 
 \item Department of Terrestrial Magnetism, Carnegie Institution for Science, 5241 Broad Branch Rd NW, Washington, DC 20015, USA
\end{affiliations}

\begin{abstract}

Exoplanet detections have revolutionized astronomy, offering new insights into solar system architecture and planet demographics. While nearly 1900 exoplanets have now been discovered and confirmed,\cite{2013PASP..125..989A} none are still in the process of formation. Transition discs, protoplanetary disks with inner clearings\cite{2011ApJ...732...42A,1989AJ.....97.1451S,2005ApJ...630L.185C}  best explained by the influence of accreting planets\cite{1999ApJ...514..344B}, are natural laboratories for the study of planet formation. Some transition discs show evidence for the presence of young planets in the form of disc asymmetries\cite{2013ApJ...775...30I,2014ApJ...783L..13P} or infrared sources detected within their clearings, as in the case of LkCa 15.\cite{2012ApJ...745....5K,2014IAUS..299..199I} Attempts to observe directly signatures of accretion onto protoplanets have hitherto proven unsuccessful.\cite{2015A&A...579A..48W} Here we report adaptive optics observations of LkCa 15 that probe within the disc clearing. With accurate source positions over multiple epochs spanning 2009 - 2015, we infer the presence of multiple companions on Keplerian orbits. We directly detect H$\alpha$ emission from the innermost companion, LkCa 15 b, evincing hot ($\sim10,000$ K) gas falling deep into the potential well of an accreting protoplanet.

\end{abstract}

We observed LkCa 15 using the high-contrast imaging technique of non-redundant masking (NRM)\cite{2000SPIE.4006..491T}, at the Large Binocular Telescope (LBT) in Ks ($\lambda_c = 2.16~ \mu$m) and L$'$ ($\lambda_c = 3.7~ \mu$m; see Extended Data Table 1). We detect two components, LkCa 15 b and c, in both bands, with consistent positions across wavelength given the uncertainties (see Table 1, Extended Data Figure 1). We detect a faint, third component, LkCa 15 d, at L$'$ only. Since d is significantly fainter than b and c, and not detected at Ks, we focus on the other two sources in the following analysis, but include discussion of the putative third companion where relevant.

We also observed LkCa 15 in H$\alpha$ ($\lambda_c = 655.8$ nm) using the Magellan Adaptive Optics System (MagAO) in Simultaneous Differential Imaging (SDI)\cite{2014ApJ...781L..30C,2003IAUS..211..275M}  mode (see Methods). We detect LkCa 15 b in these data, at a signal-to-noise of 6.4 and a position that agrees with the LBT observations (see Table 1, Extended Data Figures 2-5, Extended Data Table 2). LkCa 15 c was not detected in H$\alpha$, perhaps due to higher extinction along the line of sight or lower accretion rates at the time of the observations. Both b and c lie well within the disc clearing (Figure 1), which extends to a stellocentric radius of 56 AU.\cite{2014A&A...566A..51T}

We compare LkCa 15 b and c's positions to the infrared signal seen in 2009-2010 NRM observations.\cite{2012ApJ...745....5K} As shown in Figure 2, orbital fits (fixed to the outer disc plane: inclination $i=50^\circ$, position angle $\theta=150^\circ$)\cite{2014A&A...566A..51T} suggest distinct orbits, with b moving faster (semimajor axis, $a=14.7\pm2.1$ AU) than c ($a=18.6\pm2.5$ AU). Taking the semimajor axis uncertainties into account and requiring that these orbits be stable, b and c must have masses $ < 5 - 10~\mathrm{M_J}$\cite{1993Icar..106..247G,2003ApJ...593.1124B}, with masses $> 5~\mathrm{M_J}$ allowed only in the case of a 2:1 resonance. For completeness, we performed a series of four-body simulations to show that stable orbital solutions exist including LkCa 15 d, with three planet masses $\leq 0.5~\mathrm{M_J}$ (see Methods, Extended Data Figures 6-7) and higher masses for b and c allowed with a less massive d. 

We calculate LkCa 15 b and c's infrared fluxes, and compare them to circumplanetary accretion disc models\cite{2015ApJ...803L...4E,2015ApJ...799...16Z} and hot-start\cite{2012ApJ...745..174S} models of sub-stellar mass companions shortly after accretion has ceased (see Figure 3). From the LkCa 15 A magnitudes (2MASS Ks = 8.16\cite{2006AJ....131.1163S} and IRAC m$_{3.6}$ = 7.61\cite{2010ApJS..186..259R}), we derive fluxes of $1.4\pm0.7$ mJy at Ks and $2.5\pm1.2$ mJy at L$^\prime$ for b, and $2.3\pm1.1$ mJy at Ks and $2.5\pm1.2$ mJy at L$^\prime$ for c. These are consistent with accretion discs having inner radii $R_{in} = 2~\mathrm{R_J}$ and planet mass times accretion rate $M_p\dot{M} \sim 10^{-5}~\mathrm{M_J^2}~ \mathrm{yr^{-1}}$. However, changing $R_{in}$ affects both the total disc flux and its colour. The large uncertainties on fluxes and colours allow us to constrain $R_{in}$ only to within a factor of $\sim2$, translating to a factor of $\sim2-3$ uncertainty in $M_p\dot{M}$ (for example, a $R_{in} = 1~\mathrm{R_J}$, $M_p\dot{M} \sim 3 \times 10^{-6}~\mathrm{M_J^2~yr^{-1}}$ disc can also reproduce the observations). 

While the hot-start model shown in Figure 3 can approximately produce the Ks and L$'$ emission for b and c, the observations are best explained by an accretion disc model. The hot start model can only match a previously-established 1.55 $\mu$m upper limit on the contrast of the structure within the disc gap ($\Delta H = 7.2$ mag)\cite{2014IAUS..299..199I} if the extinction is significantly higher than inferred toward the star. Moreover, even a highly-extincted hot-start model cannot reproduce the strong emission at $4.7$ $\mu$m (contrast of $\Delta M = 3.5$).\cite{2014IAUS..299..199I} Emission from an accretion disc increases from L$'$ to M band, while the hot-start model produces little M band emission. Finally, a cooling photosphere produces no H$\alpha$ emission, firmly ruling out the hot-start model as the source of LkCa 15 b.

Since LkCa 15 b is detected at H$\alpha$, an accretion tracer\cite{2014ApJ...783L..17Z,2012A&A...548A..56R,1994ApJ...426..669H}, its nature as an accreting protoplanet is clear. LkCa 15 b's H$\alpha$ contrast, corrected for A's H$\alpha$ excess and assuming equal extinction to A ($A_R = 0.75$ mag\cite{1995ApJS..101..117K}), corresponds to a line flux of $\sim6\times10^{-5}~\mathrm{L_{\odot}}$. Assuming similar accretion luminosity ($L_{acc}$) scalings as low-mass T Tauri stars\cite{2012A&A...548A..56R,2014ApJ...781L..30C} gives $L_{acc} \sim 4\times10^{-4}~\mathrm{L_\odot}$, yielding $M_p\dot{M} \sim 3\times10^{-6}~\mathrm{M_J^2~yr^{-1}}$ for a 1.6 $\mathrm{R_J}$ planet.\cite{1998ApJ...492..323G} Previous observations showed that low-mass, accreting objects may emit a higher fraction of accretion luminosity at H$\alpha$;\cite{2014ApJ...783L..17Z} assuming similar accretion scalings as T Tauri stars may overestimate $L_{acc}$. Extinction toward b is also uncertain; while we assume equal extinction to A and b, localized extinction can alter the numbers quoted above. While the uncertainties are large, this $M_p\dot{M}$ is consistent with that estimated from the infrared fluxes. 

Previous investigators posited a single protoplanet in LkCa 15, accreting material from its co-orbital surroundings.\cite{2012ApJ...745....5K} While the semimajor axis uncertainties do not formally rule out b and c (and d, see Extended Data Figure 6) being co-orbital, physical arguments show that they cannot be gravitationally bound. The size of the previously reported emission (several AU) is larger than a Hill radius ($\sim 1.8$ AU for a 10 $\mathrm{M_J}$ planet orbiting a 1 $\mathrm{M_\odot}$ star at 10 AU), and much larger than the maximum possible size of a circumplanetary disc ($\sim1/3$ the Hill radius\cite{2011MNRAS.415..576A}). Thus the sources cannot be part of a bound, accreting system, and an alternative scenario is required to explain the observations.

We argue further that it is difficult to explain LkCa 15 b and c (and d) with an orbiting clump of gravitationally unbound dust within the disc gap, emitting thermally or in scattered light. At a distance of $\sim 10$ AU, neither LkCa 15 A nor a companion with a contrast of $\sim5$ magnitudes can heat dust sufficiently to emit at $2-4~\mu$m. Assuming isotropic single scattering, we calculate that an optically thin spherical clump of dust, perhaps resulting from a recent planetesimal collision, could produce the contrast observed at a single wavelength.  However, observing this clump before it sheared out would be \emph{a priori} unlikely, since the viscous timescale at $\sim10$ AU is just $\sim3\%$ the age of the system. 

Observations argue strongly against this explanation as well. Scattering cannot cause increasing emission from H to M band,\cite{2014IAUS..299..199I} since dust opacity decreases with increasing wavelength. Furthermore, since dust opacity is equal between H$\alpha$ and the nearby continuum, scattering signals have equal contrast in both narrowband filters. Scaling the continuum image by the LkCa 15 A H$\alpha$-to-continuum flux ratio and subtracting it from the H$\alpha$ image should only lead to an H$\alpha$ detection if scattering is not the emission mechanism. Indeed, this yields a LkCa 15 b detection with signal-to-noise of 4.8. While the Wollaston beamsplitter in MagAO's SDI mode could lead to contamination by polarized light, the visible polarized scattered light intensity at b's position is less than $\sim7\%$ the H$\alpha$ source flux.\cite{2015ApJ...808L..41T} It could not cause the H$\alpha$ detection. This leaves the multiple-planet scenario as the most natural explanation for the data.

Both the infrared and H$\alpha$ observations show that we are unambiguously witnessing planet formation in LkCa 15. The data offer evidence that giant protoplanets undergo a period of high infrared and H$\alpha$ luminosity during their accretion phase. In the near future, ALMA's sensitivity and angular resolution should enable us to detect sub-millimetre emission from circumplanetary discs.\cite{2014ApJ...788..129I} Additionally, while the LBT data published here were taken in single-aperture mode (baselines up to $\sim8$ m), non-redundant masking using the co-phased LBTI will provide 23-m baselines, allowing us to place tight constraints on the companion orbits and to resolve structure at smaller separations. Continued monitoring of accretion tracers from LkCa 15 b will probe whether the accretion is steady or stochastic. This young system provides the first opportunity to study planet formation and disc-planet interactions directly.

\section*{Methods 1: LBT Observations and Data Reduction}

We observed LkCa 15 using non-redundant masking (NRM)$^{11}$ with LBTI/LMIRCam\cite{2008SPIE.7013E..28H,2012SPIE.8446E..4FL} in December 2014 and February 2015. NRM transforms a conventional telescope into an interferometric array through the use of a pupil-plane mask, providing better PSF characterization and resolving angular scales even within $\lambda / D$.  We used LMIRCam's 12-hole mask in single-aperture mode, yielding 1.4 - 7.0 metre baselines. We broke up the observations into ``visits," each consisting of an identical set of integrations on LkCa 15 and an unresolved calibrator star (see Extended Data Table 1). We used three calibrators to lessen the possibility of contamination by a binary calibrator, and included one calibrator, GM Aur, from those observed previously at Keck.$^{8}$ We let the sky rotate throughout the observations, facilitating calibration of quasi-static speckles. At Ks and L$'$ we observed LkCa 15 at parallactic angles between $-37^\circ$ and $65^\circ$, and $-65^\circ$ and $65^\circ$, respectively.

The NRM images show the interference fringes formed by the mask, the Fourier transform of which yields complex visibilities. Sampling the complex visibilities, we calculated squared visibilities (power versus baseline) and closure phases (sums of phases around three baselines forming a triangle). Closure phases eliminate atmospheric phase errors, leaving behind the sum of the intrinsic source phases. The LBT raw closure phase scatter was $\sim3^\circ$, significantly lower than the uncertainties in previous NRM observations$^{8}$ ($\sim4^\circ$). 

For each closing triangle, we fitted a polynomial to all calibrator closure phases as a function of time. We sampled the polynomial at the time of each target observation and subtracted it from each target closure phase. We calibrated using a variety of functions; of these, polynomials up to 2nd order in time provided the lowest-scatter calibrated closure phases, with standard deviations of $1.7^\circ$ at Ks and $1.9^\circ$ at L$'$. We calibrated the squared visibilities similarly, dividing by the calibrator rather than subtracting. We calibrated the mask baselines using the observed power spectra and knowledge of the filter bandpass and plate scale\cite{2015A&A...576A.133M}.

\section*{Methods 2: LBT Image Reconstruction, Model Fitting, \& Parameter Error Estimation}

We fitted models directly to kernel phases,\cite{2010ApJ...724..464M,2013MNRAS.433.1718I} linearly independent combinations of closure phases, to search for companions. We modeled the central star as a delta function and each companion as another delta function located a distance $s$ from the star, at position angle $PA$, and with contrast $\Delta$. We sampled the synthetic complex visibilities at the same baselines and parallactic angles as the data, and performed a grid fit, using a $\Delta \chi^2$ to determine our parameter confidence intervals. Due to a known degeneracy between companion separation and contrast,\cite{2015ApJ...801...85S} brighter companions at smaller separations provide equally good fits as those fainter and farther out. We thus performed fits to individual wavelengths to verify that the positions of b and c agreed across wavelength, then calculated a best fit where the companions coincided at Ks and L$'$ (see Table 1). The model grids in this study required $\sim50,000$ CPU hours to generate, but were computed in a reasonable amount of time using the University of Arizona's El Gato supercomputer. 

We also reconstructed images from the closure phases. To produce cleaner images, we re-calibrated the closure phases toward the best-fit Ks + L$^\prime$ model using an optimized calibrator weighting technique applied in previous NRM studies.$^{8}$ This calibration is similar to the Locally Optimized Calibration of Images (LOCI)\cite{2007ApJ...660..770L} technique applied in direct imaging. Since this scheme can remove signal and underestimate errors, we applied it only to produce images (see Extended Data Figure 1), using the polynomial calibration to estimate companion parameters. As a consistency check, we reconstructed images using both the BiSpectrum Maximum Entropy Method (BSMEM)\cite{1994IAUS..158...91B} and the Monte-Carlo MArkov Chain IMager algorithm (MACIM).\cite{2006SPIE.6268E..1TI} The companion positions agree between the two algorithms, although BSMEM produces more compact emission toward each component. BSMEM has been shown to provide the most faithful image reconstruction of any available algorithms in blind tests.\cite{2006SPIE.6268E..1UL}

\section*{Methods 3: Companion Parameter Error Estimation for Previously-Published Keck Data}

Orbital fitting required parameter errors for the previously published$^8$ Keck observations and the LBT observations to be consistently estimated. The published errors for the 2009-2010 companion parameters were generated using the non-linear algorithm MPFIT.\cite{2009ASPC..411..251M} While non-linear fitters are often employed for computational expediency, the Levenberg-Marquardt algorithm can easily fall into a local minimum and underestimate parameter errors. The LBT grid $\chi^2$ surfaces show local minima for both two- and three-companion fits, rendering MPFIT unreliable unless the starting parameters were very close to the global minimum. We compared MPFIT and grid-based parameter errors for the LBT data, and found that MPFIT significantly underestimated the errors (Figure 2). 

To create a ``typical" error bar for each Keck companion, we estimated the error bar dependence on contrast using the LBT fits. Errors increased with decreasing companion flux, which we parameterized as a square root dependence. For a given Keck companion we thus scaled our LBT errors by the square root of the LBT-to-Keck flux ratio. We inflated the Keck error bars by a factor of 1.3, the ratio of the uncalibrated closure phase scatter in the Keck data ($\sim4^\circ$) to that for the LBT data ($\sim3^\circ$). We capped the separation upper limits at $3\lambda / D$, where D is Keck's telescope diameter, 10m, since the largest LBT upper limit was at nearly 3$\lambda / D$, and companions at those distances are no longer subject to the separation-contrast degeneracy.

\section*{Methods 4: MagAO Data Reduction and Analysis}

We observed LkCa 15 on November 15 and 22, 2014, as part of the Giant Accreting Protoplanet Survey (GAPplanetS), a visible-wavelength survey of bright transition discs. GAPplanetS stars are imaged simultaneously in H$\alpha$ (0.656 $\mu$m, $\Delta\lambda = 6$ nm) and the nearby stellar continuum (0.642 $\mu$m, $\Delta\lambda = 6$ nm) with the 585-actuator Magellan Adaptive Optics systemÕs SDI camera.$^{12,}$\cite{Morzinski:2014, Close:2012} The continuum channel provides a sensitive, simultaneous probe of the stellar PSF, allowing for effective removal of residual starlight and isolation of H$\alpha$ emitting sources$^{12,}$\cite{2013ApJ...775L..13F}. The observations utilized new single-substrate narrowband H$\alpha$ and continuum filters, a significant improvement over the previous VisAO SDI filters, which suffered from ghost images.$^{12}$ 

Seeing during the November 15 observations was better than the site median ($0.56 \pm 0.06''$), winds were low ($3.6 \pm 0.9$ mph), and humidity was typical of the season ($37.0\pm 2.8 \%$). Strehl ratio was low ($< 10\%$), and difficult to measure meaningfully. We characterized image quality using the stellar full-width at half-maximum (FWHM), 0.07$''$ (at 0.65 $\mu$m over 30 second integrations), a significant improvement over the seeing. We collected 316 30-second closed-AO-loop images, with a total of 156 minutes of integration time and 48.6$^\circ$ of sky rotation. We selected the 149 LkCa 15 images with the lowest residual wavefront error ($\sim50\%$), equivalent to 74.5 minutes of exposure time. This image subset had 47.6$^\circ$ of sky rotation, with the rotational space well sampled.

The November 22 data were not of sufficient quality to recover LkCa 15 b, due to lower sky rotation (27.0$^\circ$), shorter total integration (91 minutes), and shallower individual exposures (15 seconds). Injected positive planets with the same separation as LkCa 15 b were only recoverable with SNR $>3$ at contrasts $>5\times10^{-2}$ (nearly an order of magnitude brighter than the measured November 15 LkCa 15 b contrast). For this reason, we discuss only the November 15 dataset in the rest of the paper.
 
Images were first bias-subtracted, registered, and aligned via cross-correlation. The H$\alpha$ flat field image showed very little non-uniformity across the field ($<1\%$), so a flat field was not applied. We masked CCD dust spots visible in the flat field wherever they affected the image throughput by more than 2$\%$. 

We processed the aligned data using angular differential imaging (ADI\cite{Marois:2008}), comparing the ``classical'' method of using a single median point-spread function (PSF) for all images (cADI\cite{2006ApJ...641..556M}) to the Karhunen-Loeve Image Processing (KLIP\cite{2012ApJ...755L..28S}) algorithm, which calculates a least-squares optimum PSF for each image. LkCa 15 b was detected in the H$\alpha$ channel via both methods, as shown in Extended Data Figure 2. The planet was not detected in continuum with either method, so continuum images were used as a probe of PSF residuals and scattered light emission from the inner disk. Subtraction of the processed continuum images from the H$\alpha$ images (``ASDI'') left behind only true H$\alpha$ emission.$^{12}$ 

\subsection{Classical ADI reductions}
We constructed the stellar PSF by median combining images in 0.5$^\circ$ rotational bins and then median combining again to produce a PSF evenly sampled in rotational space. We subtracted the stellar PSF from the individual images, rotated them to a common on-sky orientation and combined them. Given the small separation between LkCa 15 A and b, the planet moved by only 1.5 FWHM over the course of the observations, resulting in self-subtraction and decreasing the FWHM of the processed planet PSF to $\sim4-5$ pixels in azimuth. 

\subsection{KLIP-ADI Reductions}
KLIP reductions were carried out using a well-tested custom IDL code.\cite{2014ApJ...786...32M} To optimize reduction parameters, we maximized the signal to noise of injected planets (with the same separation and contrast as LkCa 15 b) inserted after using a negative planet to erase the LkCa 15 b signal. Planets were placed at position angles distant from known artifacts, and east or west of the star to avoid the noisier north / south region of the PSF, corresponding to the wind direction during the observations.

To limit self-subtraction, the library from which KLIP builds the stellar PSF is limited to images where a planet would have rotated away from its original position. We explored the size of this exclusion region (``rotational mask") systematically through fake planet injection, and found that a $5^\circ$ mask ($\sim$1 pixel at $r=11$ pixels) produced the highest signal-to-noise recoveries of injected planets. Given the stellar FWHM of 0.07$''$, this resulted in azimuthal self-subtraction, with a processed planetary PSF of $\sim$2 pixels in azimuth.

Noise in the KLIP processed images was mostly Gaussian when images were divided into several independently-optimized radial zones, indicating efficient removal of speckles. Dividing these zones azimuthally provided no additional advantage, and the final KLIP reductions shown in Extended Data Figure 2 reflect a PSF divided into 50-pixel (0.4$''$) annuli. Removal of the median PSF radial profile for the entire image set aided significantly in attenuating the stellar halo, improving the ability of the KLIP algorithm to match residual speckles and enhancing contrast close to the star. 

\subsection{Photometry and Astrometry}

We estimated photometry and astrometry by minimizing residuals after injecting a negative planet at the location of LkCa 15 b. The cube of registered and aligned H$\alpha$ channel images was scaled by the chosen contrast value, multiplied by -1, and injected into the raw images before KLIP processing. Using the full H$\alpha$ image cube rather than its median combination simulated variability of the PSF between images. 

We generated error bars by injecting false positive planets with similar separations and contrasts to LkCa 15 b after using a negative planet to eliminate the true signal. Planets were placed at position angles away from the wind direction, and spaced by at least 75\% of the measured stellar FWHM. We computed the centroid and peak pixel using a 5 pixel aperture around each planet, and assigned the standard deviations in recovered flux and position as our 1$\sigma$ photometric and astrometric uncertainties, respectively (see Extended Data Table 2 and Extended Data Figure 3).

\subsection{Signal-to-noise of the H$\alpha$ Detection}

To create signal-to-noise ratio (SNR) maps, we calculated a radial noise profile using the standard deviation of 1-pixel-wide annuli and divided it into the raw images. In the raw maps, LkCa 15 b has SNR $\sim 3-4$. Smoothing by a Gaussian with a 2-pixel FWHM maximized the SNR of injected fake planets, so we applied this smoothing to the final science images, resulting in peak SNRs of 4.4 and 6.8 in the KLIP H$\alpha$ and ASDI images, respectively. However, directly-imaged exoplanets at small separations suffer from small number statistical effects.\cite{Mawet:2014} The underlying speckle distribution is difficult to probe given the small number of independently sampled noise regions. In an annulus at the distance of LkCa 15 b (1.3 FWHM), seven noise regions exist, leading to corrected\cite{Mawet:2014} SNRs of 4.1 and 6.4 for the H$\alpha$ and ASDI images, respectively. The ASDI detection corresponds to a false positive probability of $3\times10^{-4}$ using the Student's t-distribution with 6 degrees of freedom.

Comparing the LkCa 15 b SNR to the distribution of values in the ASDI SNR map (Extended Data Figure 4), shows that it is a clear outlier. Comparison of the peak pixel in an aperture centered on b to those in the surrounding noise apertures (Extended Data Figures 4-5) further demonstrates b's statistical significance. 

In addition to the high SNR, low false positive fraction, and the statistics presented in Extended Data Figure 4, the H$\alpha$ detection is significant because it occurs at the same location as the independent LBT detection. This further reduces the probability of a false positive detection in the MagAO data, since speckles have no preferred location. 

\subsection{Fidelity of the LkCa 15 b Detection}

Neither the existence of LkCa 15 b nor its derived parameters are dependent on our choice to include only the top $\sim50\%$ of raw images. The planet appears at the same location and with the same approximate brightness when processing all 316 images, as well as only the top 25\% of images. An excess with SNR $> 3$ appears at LkCa 15 b's location with a wide range of KLIP zone geometries and rotational masks, when any number of KL modes from 2 to 100+ are removed, and whether or not the median radial profile of the PSF is subtracted before processing. 

\subsection{Limits on LkCa 15 b SDI Continuum Flux}

We used simulated planet detections to place an upper limit on LkCa 15 b's continuum flux. We injected planets into the raw continuum channel images with a range of contrasts and at positions near LkCa 15 b. We then measured the SNR of each simulated detection to determine the confidence at which we could detect a given contrast. As above, we apply a small number statistical correction\cite{Mawet:2014} to the SNR of each recovered planet. The simulations suggest that we would have detected an excess with a corrected SNR of 3 (false positive fraction of $10^{-2}$) for a continuum source with contrast greater than $5\times10^{-3}$. Since LkCa 15 A is 1.8 times brighter at H$\alpha$ than continuum, this corresponds to an H$\alpha$-to-continuum-flux ratio lower limit of 2.7. 

\subsection{Limits on LkCa 15 c H$\alpha$ Contrast}

We established limits on the LkCa 15 c H$\alpha$ contrast using false planet injections, first using a negative planet to eliminate the LkCa 15 b signal. We injected planets with a range of contrasts at positions sampling the LBT error ellipse for LkCa 15 c. At position angles between -40$^\circ$ and -52$^\circ$, several $2-2.5~\sigma$ peaks near c's location boost the SNRs for recovered planets. Here, we can detect contrasts down to $2\times 10^{-3}$ with corrected\cite{Mawet:2014} SNRs of 3 (false positive fraction of $2 \times 10^{-2}$). Position angles greater than -40$^\circ$ approach a noisier region of the PSF, leading to decreased sensitivity; here contrasts of $6\times10^{-3}$ are required for adjusted signal-to-noise ratios of 3. We cannot reject or confirm accretion onto LkCa 15 c below $6\times10^{-3}$ contrast ($\Delta H\alpha = 5.6$) with the current dataset. This improves upon previous spectro-astrometric observations, which yielded a contrast limit of $\Delta H\alpha = 3.4$ for a position angle near LkCa 15 c.$^{10}$

\section*{Methods 5: Stability Analysis with LkCa 15 d}

We ran a series of orbit integrations to demonstrate that stable solutions exist for b, c, and d at separations within the 1$\sigma$ semimajor axis error bars (see Extended Data Figures 6-7). We used the publicly available Swifter package\cite{2013ascl.soft03001L} - specifically, the symplectic integrator, SyMBA\cite{1998AJ....116.2067D}, which switches to a Burlisch-Stoer algorithm for planetary close approaches. We also ran comparison integrations with the Gauss Radau 15th order integrator and found comparable results, with minimum energy conservation of 1 part in $10^{7}$ over a 10 Myr integration.

We required all orbits to be nearly co-planar, with a random scatter $< 1^\circ$, and assigned each planet a random eccentricity below $10^{-4}$. To assess stability we integrated three different random phase combinations for 10 Myr. We found stable three body solutions out to $1-2$ Myr with semi-major axes of $a_b =12.7$ AU, $a_c = 18.6$ AU, $a_d = 24.7$ AU. To ensure stability out to $10$ Myr with orbits in the $1\sigma$ errors requires that all planets be $\leq0.5~M_J$. A wider range of orbits are allowed if d's mass is decreased further. These constraints are in line with previous large numerical studies of equally spaced (in $R_{\rm H,m}$), equal-mass planets.\cite{2007MNRAS.382.1823F} Note for planets b and c, there are possible resonant configurations within the predicted period ranges, which would admit somewhat higher masses. 

\bibliography{manuscript_2}

\begin{thebibliography}{10}
\expandafter\ifx\csname url\endcsname\relax
  \def\url#1{\texttt{#1}}\fi
\expandafter\ifx\csname urlprefix\endcsname\relax\def\urlprefix{URL }\fi
\providecommand{\bibinfo}[2]{#2}
\providecommand{\eprint}[2][]{\url{#2}}

\bibitem{2013PASP..125..989A}
\bibinfo{author}{{Akeson}, R.~L.} \emph{et~al.}
\newblock \bibinfo{title}{{The NASA Exoplanet Archive: Data and Tools for
  Exoplanet Research}}.
\newblock \emph{\bibinfo{journal}{\pasp}} \textbf{\bibinfo{volume}{125}},
  \bibinfo{pages}{989--999} (\bibinfo{year}{2013}).
\newblock \eprint{1307.2944}.

\bibitem{2011ApJ...732...42A}
\bibinfo{author}{{Andrews}, S.~M.} \emph{et~al.}
\newblock \bibinfo{title}{{Resolved Images of Large Cavities in Protoplanetary
  Transition Disks}}.
\newblock \emph{\bibinfo{journal}{\apj}} \textbf{\bibinfo{volume}{732}},
  \bibinfo{pages}{42--66} (\bibinfo{year}{2011}).

\bibitem{1989AJ.....97.1451S}
\bibinfo{author}{{Strom}, K.~M.}, \bibinfo{author}{{Strom}, S.~E.},
  \bibinfo{author}{{Edwards}, S.}, \bibinfo{author}{{Cabrit}, S.} \&
  \bibinfo{author}{{Skrutskie}, M.~F.}
\newblock \bibinfo{title}{{Circumstellar material associated with solar-type
  pre-main-sequence stars - A possible constraint on the timescale for planet
  building}}.
\newblock \emph{\bibinfo{journal}{\aj}} \textbf{\bibinfo{volume}{97}},
  \bibinfo{pages}{1451--1470} (\bibinfo{year}{1989}).

\bibitem{2005ApJ...630L.185C}
\bibinfo{author}{{Calvet}, N.} \emph{et~al.}
\newblock \bibinfo{title}{{Disks in Transition in the Taurus Population:
  Spitzer IRS Spectra of GM Aurigae and DM Tauri}}.
\newblock \emph{\bibinfo{journal}{\apjl}} \textbf{\bibinfo{volume}{630}},
  \bibinfo{pages}{L185--L188} (\bibinfo{year}{2005}).

\bibitem{1999ApJ...514..344B}
\bibinfo{author}{{Bryden}, G.}, \bibinfo{author}{{Chen}, X.},
  \bibinfo{author}{{Lin}, D.~N.~C.}, \bibinfo{author}{{Nelson}, R.~P.} \&
  \bibinfo{author}{{Papaloizou}, J.~C.~B.}
\newblock \bibinfo{title}{{Tidally Induced Gap Formation in Protostellar Disks:
  Gap Clearing and Suppression of Protoplanetary Growth}}.
\newblock \emph{\bibinfo{journal}{\apj}} \textbf{\bibinfo{volume}{514}},
  \bibinfo{pages}{344--367} (\bibinfo{year}{1999}).

\bibitem{2013ApJ...775...30I}
\bibinfo{author}{{Isella}, A.} \emph{et~al.}
\newblock \bibinfo{title}{{An Azimuthal Asymmetry in the LkH{$\alpha$} 330
  Disk}}.
\newblock \emph{\bibinfo{journal}{\apj}} \textbf{\bibinfo{volume}{775}},
  \bibinfo{pages}{30--40} (\bibinfo{year}{2013}).

\bibitem{2014ApJ...783L..13P}
\bibinfo{author}{{P{\'e}rez}, L.~M.}, \bibinfo{author}{{Isella}, A.},
  \bibinfo{author}{{Carpenter}, J.~M.} \& \bibinfo{author}{{Chandler}, C.~J.}
\newblock \bibinfo{title}{{Large-scale Asymmetries in the Transitional Disks of
  SAO 206462 and SR 21}}.
\newblock \emph{\bibinfo{journal}{\apjl}} \textbf{\bibinfo{volume}{783}},
  \bibinfo{pages}{L13--L18} (\bibinfo{year}{2014}).

\bibitem{2012ApJ...745....5K}
\bibinfo{author}{{Kraus}, A.~L.} \& \bibinfo{author}{{Ireland}, M.~J.}
\newblock \bibinfo{title}{{LkCa 15: A Young Exoplanet Caught at Formation?}}
\newblock \emph{\bibinfo{journal}{\apj}} \textbf{\bibinfo{volume}{745}},
  \bibinfo{pages}{5--16} (\bibinfo{year}{2012}).

\bibitem{2014IAUS..299..199I}
\bibinfo{author}{{Ireland}, M.~J.} \& \bibinfo{author}{{Kraus}, A.~L.}
\newblock \bibinfo{title}{{Orbital Motion and Multi-Wavelength Monitoring of
  LkCa15 b}}.
\newblock In \bibinfo{editor}{{Booth}, M.}, \bibinfo{editor}{{Matthews}, B.~C.}
  \& \bibinfo{editor}{{Graham}, J.~R.} (eds.) \emph{\bibinfo{booktitle}{IAU
  Symposium}}, vol. \bibinfo{volume}{299} of \emph{\bibinfo{series}{IAU
  Symposium}}, \bibinfo{pages}{199--203} (\bibinfo{year}{2014}).

\bibitem{2015A&A...579A..48W}
\bibinfo{author}{{Whelan}, E.~T.} \emph{et~al.}
\newblock \bibinfo{title}{{Spectro-astrometry of LkCa 15 with X-Shooter:
  Searching for emission from LkCa 15b}}.
\newblock \emph{\bibinfo{journal}{\aap}} \textbf{\bibinfo{volume}{579}},
  \bibinfo{pages}{A48} (\bibinfo{year}{2015}).

\bibitem{2000SPIE.4006..491T}
\bibinfo{author}{{Tuthill}, P.~G.}, \bibinfo{author}{{Monnier}, J.~D.} \&
  \bibinfo{author}{{Danchi}, W.~C.}
\newblock \bibinfo{title}{{Aperture masking interferometry on the Keck I
  Telescope: new results from the diffraction limit}}.
\newblock In \bibinfo{editor}{{L{\'e}na}, P.} \&
  \bibinfo{editor}{{Quirrenbach}, A.} (eds.)
  \emph{\bibinfo{booktitle}{Interferometry in Optical Astronomy}}, vol.
  \bibinfo{volume}{4006} of \emph{\bibinfo{series}{Society of Photo-Optical
  Instrumentation Engineers (SPIE) Conference Series}},
  \bibinfo{pages}{491--498} (\bibinfo{year}{2000}).

\bibitem{2014ApJ...781L..30C}
\bibinfo{author}{{Close}, L.~M.} \emph{et~al.}
\newblock \bibinfo{title}{{Discovery of H{$\alpha$} Emission from the Close
  Companion inside the Gap of Transitional Disk HD 142527}}.
\newblock \emph{\bibinfo{journal}{\apjl}} \textbf{\bibinfo{volume}{781}},
  \bibinfo{pages}{L30--L34} (\bibinfo{year}{2014}).

\bibitem{2003IAUS..211..275M}
\bibinfo{author}{{Marois}, C.}, \bibinfo{author}{{Nadeau}, D.},
  \bibinfo{author}{{Doyon}, R.}, \bibinfo{author}{{Racine}, R.} \&
  \bibinfo{author}{{Walker}, G.~A.~H.}
\newblock \bibinfo{title}{{Differential Simultaneous Imaging and Faint
  Companions: TRIDENT First Results from CFHT}}.
\newblock In \bibinfo{editor}{{Mart{\'{\i}}n}, E.} (ed.)
  \emph{\bibinfo{booktitle}{Brown Dwarfs}}, vol. \bibinfo{volume}{211} of
  \emph{\bibinfo{series}{IAU Symposium}}, \bibinfo{pages}{275--278}
  (\bibinfo{year}{2003}).

\bibitem{2014A&A...566A..51T}
\bibinfo{author}{{Thalmann}, C.} \emph{et~al.}
\newblock \bibinfo{title}{{The architecture of the LkCa 15 transitional disk
  revealed by high-contrast imaging}}.
\newblock \emph{\bibinfo{journal}{\aap}} \textbf{\bibinfo{volume}{566}},
  \bibinfo{pages}{A51} (\bibinfo{year}{2014}).

\bibitem{1993Icar..106..247G}
\bibinfo{author}{{Gladman}, B.}
\newblock \bibinfo{title}{{Dynamics of systems of two close planets}}.
\newblock \emph{\bibinfo{journal}{\icarus}} \textbf{\bibinfo{volume}{106}},
  \bibinfo{pages}{247--263} (\bibinfo{year}{1993}).

\bibitem{2003ApJ...593.1124B}
\bibinfo{author}{{Beaug{\'e}}, C.}, \bibinfo{author}{{Ferraz-Mello}, S.} \&
  \bibinfo{author}{{Michtchenko}, T.~A.}
\newblock \bibinfo{title}{{Extrasolar Planets in Mean-Motion Resonance: Apses
  Alignment and Asymmetric Stationary Solutions}}.
\newblock \emph{\bibinfo{journal}{\apj}} \textbf{\bibinfo{volume}{593}},
  \bibinfo{pages}{1124--1133} (\bibinfo{year}{2003}).

\bibitem{2015ApJ...803L...4E}
\bibinfo{author}{{Eisner}, J.~A.}
\newblock \bibinfo{title}{{Spectral Energy Distributions of Accreting
  Protoplanets}}.
\newblock \emph{\bibinfo{journal}{\apjl}} \textbf{\bibinfo{volume}{803}},
  \bibinfo{pages}{L4--L8} (\bibinfo{year}{2015}).

\bibitem{2015ApJ...799...16Z}
\bibinfo{author}{{Zhu}, Z.}
\newblock \bibinfo{title}{{Accreting Circumplanetary Disks: Observational
  Signatures}}.
\newblock \emph{\bibinfo{journal}{\apj}} \textbf{\bibinfo{volume}{799}},
  \bibinfo{pages}{16--24} (\bibinfo{year}{2015}).

\bibitem{2012ApJ...745..174S}
\bibinfo{author}{{Spiegel}, D.~S.} \& \bibinfo{author}{{Burrows}, A.}
\newblock \bibinfo{title}{{Spectral and Photometric Diagnostics of Giant Planet
  Formation Scenarios}}.
\newblock \emph{\bibinfo{journal}{\apj}} \textbf{\bibinfo{volume}{745}},
  \bibinfo{pages}{174--188} (\bibinfo{year}{2012}).

\bibitem{2006AJ....131.1163S}
\bibinfo{author}{{Skrutskie}, M.~F.} \emph{et~al.}
\newblock \bibinfo{title}{{The Two Micron All Sky Survey (2MASS)}}.
\newblock \emph{\bibinfo{journal}{\aj}} \textbf{\bibinfo{volume}{131}},
  \bibinfo{pages}{1163--1183} (\bibinfo{year}{2006}).

\bibitem{2010ApJS..186..259R}
\bibinfo{author}{{Rebull}, L.~M.} \emph{et~al.}
\newblock \bibinfo{title}{{The Taurus Spitzer Survey: New Candidate Taurus
  Members Selected Using Sensitive Mid-Infrared Photometry}}.
\newblock \emph{\bibinfo{journal}{\apjs}} \textbf{\bibinfo{volume}{186}},
  \bibinfo{pages}{259--307} (\bibinfo{year}{2010}).

\bibitem{2014ApJ...783L..17Z}
\bibinfo{author}{{Zhou}, Y.}, \bibinfo{author}{{Herczeg}, G.~J.},
  \bibinfo{author}{{Kraus}, A.~L.}, \bibinfo{author}{{Metchev}, S.} \&
  \bibinfo{author}{{Cruz}, K.~L.}
\newblock \bibinfo{title}{{Accretion onto Planetary Mass Companions of Low-mass
  Young Stars}}.
\newblock \emph{\bibinfo{journal}{\apjl}} \textbf{\bibinfo{volume}{783}},
  \bibinfo{pages}{L17--L22} (\bibinfo{year}{2014}).

\bibitem{2012A&A...548A..56R}
\bibinfo{author}{{Rigliaco}, E.} \emph{et~al.}
\newblock \bibinfo{title}{{X-shooter spectroscopy of young stellar objects. I.
  Mass accretion rates of low-mass T Tauri stars in {$\sigma$} Orionis}}.
\newblock \emph{\bibinfo{journal}{\aap}} \textbf{\bibinfo{volume}{548}},
  \bibinfo{pages}{A56} (\bibinfo{year}{2012}).

\bibitem{1994ApJ...426..669H}
\bibinfo{author}{{Hartmann}, L.}, \bibinfo{author}{{Hewett}, R.} \&
  \bibinfo{author}{{Calvet}, N.}
\newblock \bibinfo{title}{{Magnetospheric accretion models for T Tauri stars.
  1: Balmer line profiles without rotation}}.
\newblock \emph{\bibinfo{journal}{\apj}} \textbf{\bibinfo{volume}{426}},
  \bibinfo{pages}{669--687} (\bibinfo{year}{1994}).

\bibitem{1995ApJS..101..117K}
\bibinfo{author}{{Kenyon}, S.~J.} \& \bibinfo{author}{{Hartmann}, L.}
\newblock \bibinfo{title}{{Pre-Main-Sequence Evolution in the Taurus-Auriga
  Molecular Cloud}}.
\newblock \emph{\bibinfo{journal}{\apjs}} \textbf{\bibinfo{volume}{101}},
  \bibinfo{pages}{117--171} (\bibinfo{year}{1995}).

\bibitem{1998ApJ...492..323G}
\bibinfo{author}{{Gullbring}, E.}, \bibinfo{author}{{Hartmann}, L.},
  \bibinfo{author}{{Brice{\~n}o}, C.} \& \bibinfo{author}{{Calvet}, N.}
\newblock \bibinfo{title}{{Disk Accretion Rates for T Tauri Stars}}.
\newblock \emph{\bibinfo{journal}{\apj}} \textbf{\bibinfo{volume}{492}},
  \bibinfo{pages}{323--341} (\bibinfo{year}{1998}).

\bibitem{2011MNRAS.415..576A}
\bibinfo{author}{{Ayliffe}, B.~A.} \& \bibinfo{author}{{Bate}, M.~R.}
\newblock \bibinfo{title}{{Migration of protoplanets with surfaces through
  discs with steep temperature gradients}}.
\newblock \emph{\bibinfo{journal}{\mnras}} \textbf{\bibinfo{volume}{415}},
  \bibinfo{pages}{576--586} (\bibinfo{year}{2011}).

\bibitem{2015ApJ...808L..41T}
\bibinfo{author}{{Thalmann}, C.} \emph{et~al.}
\newblock \bibinfo{title}{{Optical Imaging Polarimetry of the LkCa 15
  Protoplanetary Disk with SPHERE ZIMPOL}}.
\newblock \emph{\bibinfo{journal}{\apjl}} \textbf{\bibinfo{volume}{808}},
  \bibinfo{pages}{L41--L47} (\bibinfo{year}{2015}).

\bibitem{2014ApJ...788..129I}
\bibinfo{author}{{Isella}, A.}, \bibinfo{author}{{Chandler}, C.~J.},
  \bibinfo{author}{{Carpenter}, J.~M.}, \bibinfo{author}{{P{\'e}rez}, L.~M.} \&
  \bibinfo{author}{{Ricci}, L.}
\newblock \bibinfo{title}{{Searching for Circumplanetary Disks around LkCa
  15}}.
\newblock \emph{\bibinfo{journal}{\apj}} \textbf{\bibinfo{volume}{788}},
  \bibinfo{pages}{129--135} (\bibinfo{year}{2014}).

\bibitem{2008SPIE.7013E..28H}
\bibinfo{author}{{Hinz}, P.~M.} \emph{et~al.}
\newblock \bibinfo{title}{{Status of the LBT interferometer}}.
\newblock In \emph{\bibinfo{booktitle}{Society of Photo-Optical Instrumentation
  Engineers (SPIE) Conference Series}}, vol. \bibinfo{volume}{7013} of
  \emph{\bibinfo{series}{Society of Photo-Optical Instrumentation Engineers
  (SPIE) Conference Series}}, \bibinfo{pages}{28--36} (\bibinfo{year}{2008}).

\bibitem{2012SPIE.8446E..4FL}
\bibinfo{author}{{Leisenring}, J.~M.} \emph{et~al.}
\newblock \bibinfo{title}{{On-sky operations and performance of LMIRcam at the
  Large Binocular Telescope}}.
\newblock In \emph{\bibinfo{booktitle}{Society of Photo-Optical Instrumentation
  Engineers (SPIE) Conference Series}}, vol. \bibinfo{volume}{8446} of
  \emph{\bibinfo{series}{Society of Photo-Optical Instrumentation Engineers
  (SPIE) Conference Series}}, \bibinfo{pages}{4--19} (\bibinfo{year}{2012}).

\bibitem{2015A&A...576A.133M}
\bibinfo{author}{{Maire}, A.-L.} \emph{et~al.}
\newblock \bibinfo{title}{{The LEECH Exoplanet Imaging Survey. Further
  constraints on the planet architecture of the HR 8799 system}}.
\newblock \emph{\bibinfo{journal}{\aap}} \textbf{\bibinfo{volume}{576}},
  \bibinfo{pages}{A133} (\bibinfo{year}{2015}).

\bibitem{2010ApJ...724..464M}
\bibinfo{author}{{Martinache}, F.}
\newblock \bibinfo{title}{{Kernel Phase in Fizeau Interferometry}}.
\newblock \emph{\bibinfo{journal}{\apj}} \textbf{\bibinfo{volume}{724}},
  \bibinfo{pages}{464--469} (\bibinfo{year}{2010}).

\bibitem{2013MNRAS.433.1718I}
\bibinfo{author}{{Ireland}, M.~J.}
\newblock \bibinfo{title}{{Phase errors in diffraction-limited imaging:
  contrast limits for sparse aperture masking}}.
\newblock \emph{\bibinfo{journal}{\mnras}} \textbf{\bibinfo{volume}{433}},
  \bibinfo{pages}{1718--1728} (\bibinfo{year}{2013}).

\bibitem{2015ApJ...801...85S}
\bibinfo{author}{{Sallum}, S.} \emph{et~al.}
\newblock \bibinfo{title}{{New Spatially Resolved Observations of the T Cha
  Transition Disk and Constraints on the Previously Claimed Substellar
  Companion}}.
\newblock \emph{\bibinfo{journal}{\apj}} \textbf{\bibinfo{volume}{801}},
  \bibinfo{pages}{85--107} (\bibinfo{year}{2015}).

\bibitem{2007ApJ...660..770L}
\bibinfo{author}{{Lafreni{\`e}re}, D.}, \bibinfo{author}{{Marois}, C.},
  \bibinfo{author}{{Doyon}, R.}, \bibinfo{author}{{Nadeau}, D.} \&
  \bibinfo{author}{{Artigau}, {\'E}.}
\newblock \bibinfo{title}{{A New Algorithm for Point-Spread Function
  Subtraction in High-Contrast Imaging: A Demonstration with Angular
  Differential Imaging}}.
\newblock \emph{\bibinfo{journal}{\apj}} \textbf{\bibinfo{volume}{660}},
  \bibinfo{pages}{770--780} (\bibinfo{year}{2007}).
\newblock \eprint{astro-ph/0702697}.

\bibitem{1994IAUS..158...91B}
\bibinfo{author}{{Buscher}, D.~F.}
\newblock \bibinfo{title}{{Direct maximum-entropy image reconstruction from the
  bispectrum}}.
\newblock In \bibinfo{editor}{{Robertson}, J.~G.} \& \bibinfo{editor}{{Tango},
  W.~J.} (eds.) \emph{\bibinfo{booktitle}{Very High Angular Resolution
  Imaging}}, vol. \bibinfo{volume}{158} of \emph{\bibinfo{series}{IAU
  Symposium}}, \bibinfo{pages}{91--93} (\bibinfo{year}{1994}).

\bibitem{2006SPIE.6268E..1TI}
\bibinfo{author}{{Ireland}, M.~J.}, \bibinfo{author}{{Monnier}, J.~D.} \&
  \bibinfo{author}{{Thureau}, N.}
\newblock \bibinfo{title}{{Monte-Carlo imaging for optical interferometry}}.
\newblock In \emph{\bibinfo{booktitle}{Society of Photo-Optical Instrumentation
  Engineers (SPIE) Conference Series}}, vol. \bibinfo{volume}{6268} of
  \emph{\bibinfo{series}{Society of Photo-Optical Instrumentation Engineers
  (SPIE) Conference Series}}, \bibinfo{pages}{1T1--1T8} (\bibinfo{year}{2006}).

\bibitem{2006SPIE.6268E..1UL}
\bibinfo{author}{{Lawson}, P.~R.} \emph{et~al.}
\newblock \bibinfo{title}{{2006 interferometry imaging beauty contest}}.
\newblock In \emph{\bibinfo{booktitle}{Society of Photo-Optical Instrumentation
  Engineers (SPIE) Conference Series}}, vol. \bibinfo{volume}{6268} of
  \emph{\bibinfo{series}{Society of Photo-Optical Instrumentation Engineers
  (SPIE) Conference Series}}, \bibinfo{pages}{1U1--1U12}
  (\bibinfo{year}{2006}).

\bibitem{2009ASPC..411..251M}
\bibinfo{author}{{Markwardt}, C.~B.}
\newblock \bibinfo{title}{{Non-linear Least-squares Fitting in IDL with
  MPFIT}}.
\newblock In \bibinfo{editor}{{Bohlender}, D.~A.}, \bibinfo{editor}{{Durand},
  D.} \& \bibinfo{editor}{{Dowler}, P.} (eds.)
  \emph{\bibinfo{booktitle}{Astronomical Data Analysis Software and Systems
  XVIII}}, vol. \bibinfo{volume}{411} of \emph{\bibinfo{series}{Astronomical
  Society of the Pacific Conference Series}}, \bibinfo{pages}{251}
  (\bibinfo{year}{2009}).
\newblock \eprint{0902.2850}.

\bibitem{Morzinski:2014}
\bibinfo{author}{{Morzinski}, K.} \emph{et~al.}
\newblock \bibinfo{title}{{MagAO: Status and on-sky performance of the Magellan
  adaptive optics system}}.
\newblock In \emph{\bibinfo{booktitle}{Society of Photo-Optical Instrumentation
  Engineers (SPIE) Conference Series}}, vol. \bibinfo{volume}{914804} of
  \emph{\bibinfo{series}{Society of Photo-Optical Instrumentation Engineers
  (SPIE) Conference Series}}, \bibinfo{pages}{1--13} (\bibinfo{year}{2014}).

\bibitem{Close:2012}
\bibinfo{author}{{Close}, L.} \emph{et~al.}
\newblock \bibinfo{title}{{First closed-loop visible AO test results for the
  advanced adaptive secondary AO system for the Magellan Telescope: MagAO's
  performance and status}}.
\newblock In \emph{\bibinfo{booktitle}{Society of Photo-Optical Instrumentation
  Engineers (SPIE) Conference Series}}, vol. \bibinfo{volume}{8447} of
  \emph{\bibinfo{series}{Society of Photo-Optical Instrumentation Engineers
  (SPIE) Conference Series}}, \bibinfo{pages}{0X1--0X16}
  (\bibinfo{year}{2012}).

\bibitem{2013ApJ...775L..13F}
\bibinfo{author}{{Follette}, K.~B.} \emph{et~al.}
\newblock \bibinfo{title}{{The First Circumstellar Disk Imaged in Silhouette at
  Visible Wavelengths with Adaptive Optics: MagAO Imaging of Orion 218-354}}.
\newblock \emph{\bibinfo{journal}{\apjl}} \textbf{\bibinfo{volume}{775}},
  \bibinfo{pages}{L13--L17} (\bibinfo{year}{2013}).

\bibitem{Marois:2008}
\bibinfo{author}{{Marois}, C.} \emph{et~al.}
\newblock \bibinfo{title}{{Direct Imaging of Multiple Planets Orbiting the Star
  HR 8799}}.
\newblock \emph{\bibinfo{journal}{Science}} \textbf{\bibinfo{volume}{322}},
  \bibinfo{pages}{1348--1352} (\bibinfo{year}{2008}).

\bibitem{2006ApJ...641..556M}
\bibinfo{author}{{Marois}, C.}, \bibinfo{author}{{Lafreni{\`e}re}, D.},
  \bibinfo{author}{{Doyon}, R.}, \bibinfo{author}{{Macintosh}, B.} \&
  \bibinfo{author}{{Nadeau}, D.}
\newblock \bibinfo{title}{{Angular Differential Imaging: A Powerful
  High-Contrast Imaging Technique}}.
\newblock \emph{\bibinfo{journal}{\apj}} \textbf{\bibinfo{volume}{641}},
  \bibinfo{pages}{556--564} (\bibinfo{year}{2006}).

\bibitem{2012ApJ...755L..28S}
\bibinfo{author}{{Soummer}, R.}, \bibinfo{author}{{Pueyo}, L.} \&
  \bibinfo{author}{{Larkin}, J.}
\newblock \bibinfo{title}{{Detection and Characterization of Exoplanets and
  Disks Using Projections on Karhunen-Lo{\`e}ve Eigenimages}}.
\newblock \emph{\bibinfo{journal}{\apjl}} \textbf{\bibinfo{volume}{755}},
  \bibinfo{pages}{L28--L32} (\bibinfo{year}{2012}).

\bibitem{2014ApJ...786...32M}
\bibinfo{author}{{Males}, J.~R.} \emph{et~al.}
\newblock \bibinfo{title}{{Magellan Adaptive Optics First-light Observations of
  the Exoplanet {$\beta$} Pic B. I. Direct Imaging in the Far-red Optical with
  MagAO+VisAO and in the Near-ir with NICI}}.
\newblock \emph{\bibinfo{journal}{\apj}} \textbf{\bibinfo{volume}{786}},
  \bibinfo{pages}{32--53} (\bibinfo{year}{2014}).

\bibitem{Mawet:2014}
\bibinfo{author}{{Mawet}, D.} \emph{et~al.}
\newblock \bibinfo{title}{{Fundamental Limitations of High Contrast Imaging Set
  by Small Sample Statistics}}.
\newblock \emph{\bibinfo{journal}{\apj}} \textbf{\bibinfo{volume}{792}},
  \bibinfo{pages}{97--107} (\bibinfo{year}{2014}).

\bibitem{2013ascl.soft03001L}
\bibinfo{author}{{Levison}, H.~F.} \& \bibinfo{author}{{Duncan}, M.~J.}
\newblock \bibinfo{title}{{SWIFT: A solar system integration software
  package}}.
\newblock \bibinfo{howpublished}{Astrophysics Source Code Library}
  (\bibinfo{year}{2013}).
\newblock \eprint{1303.001}.

\bibitem{1998AJ....116.2067D}
\bibinfo{author}{{Duncan}, M.~J.}, \bibinfo{author}{{Levison}, H.~F.} \&
  \bibinfo{author}{{Lee}, M.~H.}
\newblock \bibinfo{title}{{A Multiple Time Step Symplectic Algorithm for
  Integrating Close Encounters}}.
\newblock \emph{\bibinfo{journal}{\aj}} \textbf{\bibinfo{volume}{116}},
  \bibinfo{pages}{2067--2077} (\bibinfo{year}{1998}).

\bibitem{2007MNRAS.382.1823F}
\bibinfo{author}{{Faber}, P.} \& \bibinfo{author}{{Quillen}, A.~C.}
\newblock \bibinfo{title}{{The total number of giant planets in debris discs
  with central clearings}}.
\newblock \emph{\bibinfo{journal}{\mnras}} \textbf{\bibinfo{volume}{382}},
  \bibinfo{pages}{1823--1828} (\bibinfo{year}{2007}).

\end{thebibliography}

\begin{addendum}

\item This work was supported by NSF AAG grant $\#$1211329 and NASA OSS grant NNX14AD20G. This material is based upon work supported by the National Science Foundation under Grant No. 1228509. This work was performed in part under contract with the California Institute of Technology (Caltech) funded by NASA through the Sagan Fellowship Program executed by the NASA Exoplanet Science Institute. This material is based upon work supported by the National Science Foundation Graduate Research Fellowship under Grant No. DGE-1143953. Any opinion, findings, and conclusions or recommendations expressed in this material are those of the authors(s) and do not necessarily reflect the views of the National Science Foundation. 
 \item[Competing Interests] The authors declare that they have no
competing financial interests.
 \item[Correspondence] Correspondence and requests for materials
should be addressed to Steph Sallum.~(email: ssallum@email.arizona.edu).
\end{addendum}

\pagebreak

\renewcommand{\figurename}{\textbf{Table}}

\begin{figure*}
\includegraphics[width=\textwidth]{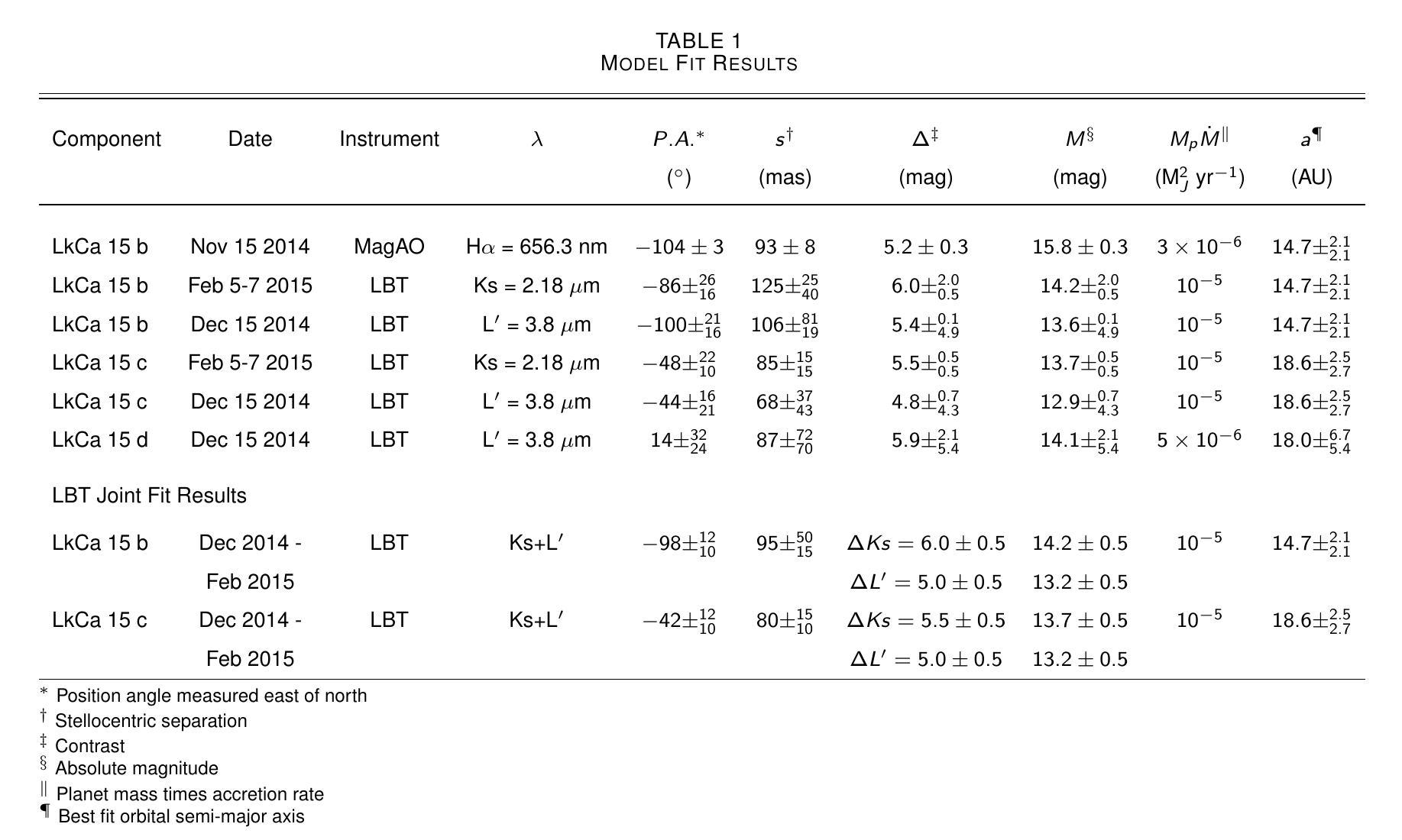}
\end{figure*}

\renewcommand{\figurename}{\textbf{Figure}}
\setcounter{figure}{0}

\begin{figure}
\centering
\includegraphics[width=\textwidth]{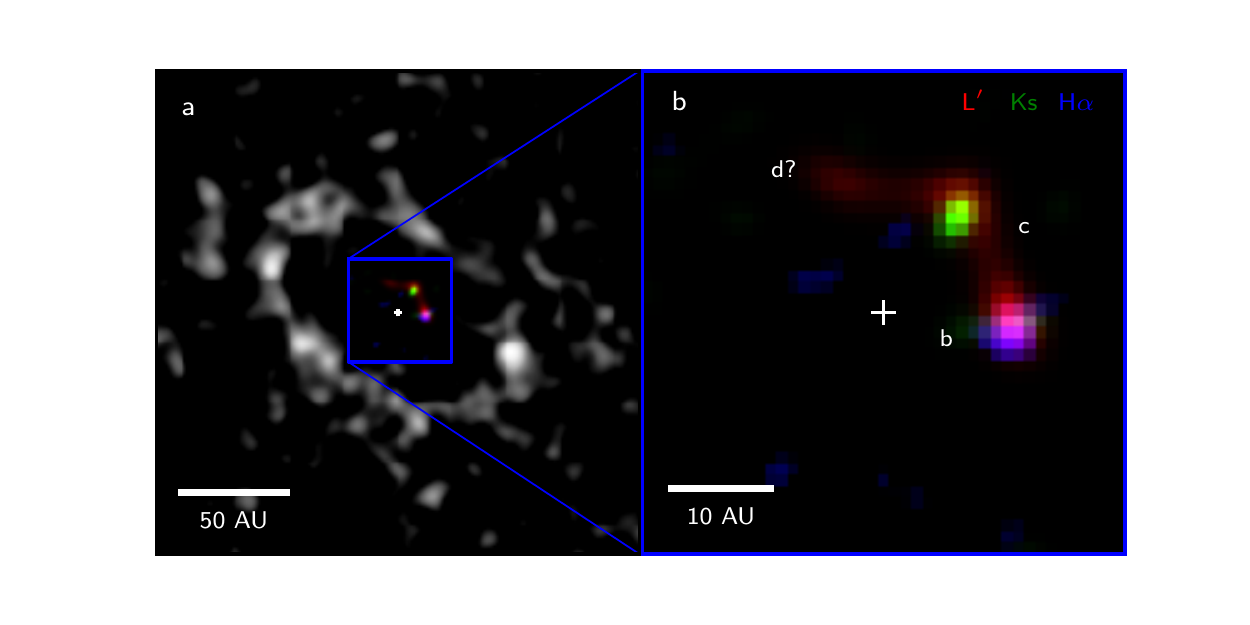}
\caption{\textbf{Composite H$\alpha$, Ks, and L$'$ image.} \textbf{a)} The coloured image shows H$\alpha$ (blue), Ks (green), and L$'$ (red) detections at the same scale as VLA millimetre observations\cite{2014ApJ...788..129I} (greyscale). \textbf{b)} Zoomed in composite image of LBT and Magellan observations, with b, c, and d marked.}
\label{fig-comp}
\end{figure}

\begin{figure}
\centering
\includegraphics[width=\textwidth]{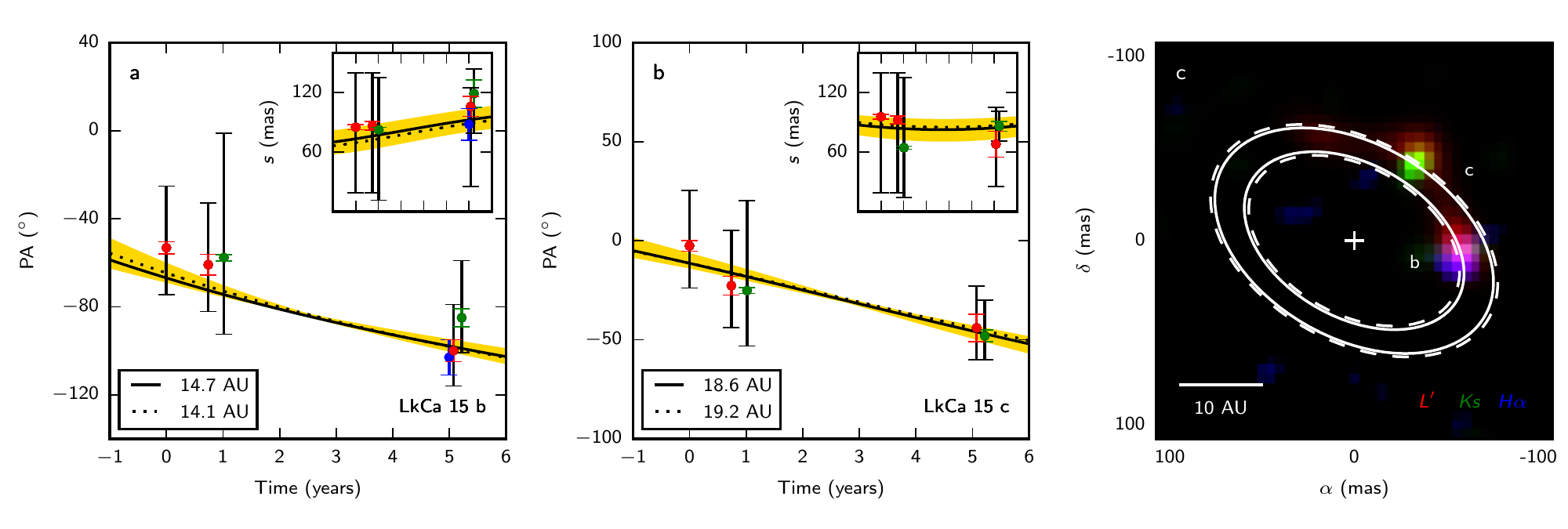}
\caption{\textbf{Position evolution.}  \textbf{a)} LkCa 15 b position angle and separation (inset) evolution, showing H$\alpha$ (blue), Ks (green), and L$'$ (red). The earliest three points indicate previous observations\cite{2012ApJ...745....5K}; others show fits to our data. Coloured and black 1$\sigma$ error bars are from a non-linear algorithm and a grid, respectively (see Methods). The yellow shading spans the $1\sigma$ allowed parameters from orbital fitting. Solid and dotted curves show stable orbits for 0.5 M$_J$ and 1.0 M$_J$ planets, respectively. \textbf{b)} Same as \textbf{a}, for LkCa 15 c. \textbf{c)} Stable orbits for 0.5 M$_J$ (solid) and 1.0 M$_J$ (dotted) planets.}
\label{fig-orb}
\end{figure}

\begin{figure}
\centering
\includegraphics[width=0.75\textwidth]{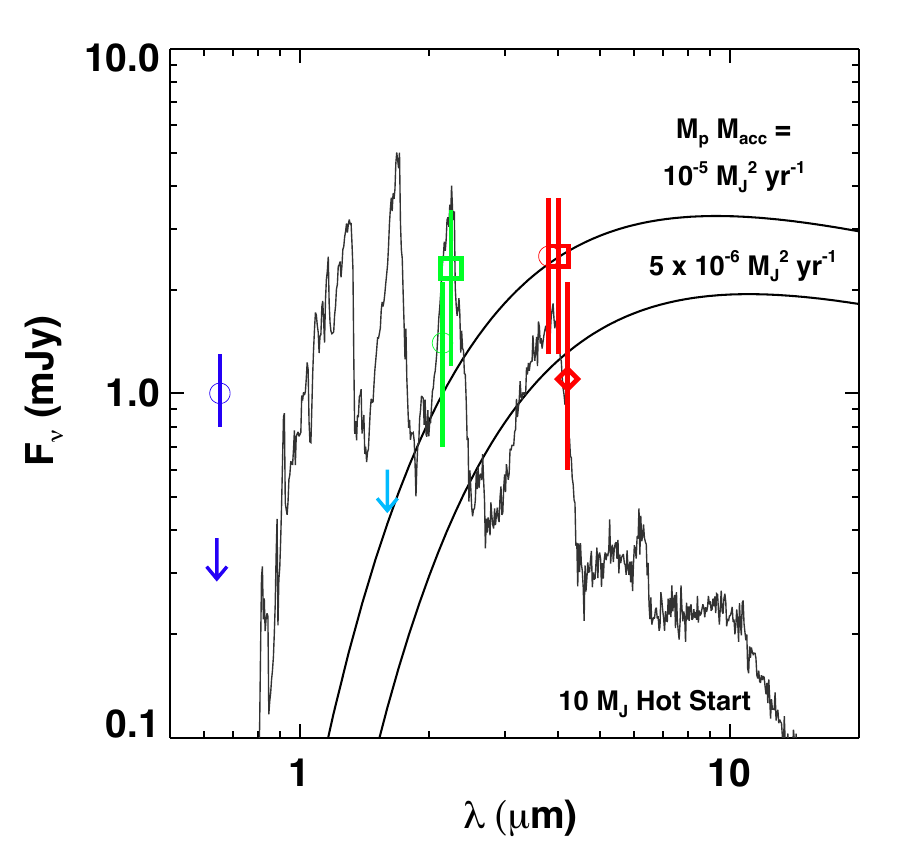}
\caption{\textbf{Spectral energy distributions.} Symbols indicate fluxes for LkCa 15 b (circles), c (squares), and d (diamonds), showing H$\alpha$ (dark blue), Ks (green), and L$^\prime$ (red). The light and dark blue arrows mark previously-published H-band\cite{2014IAUS..299..199I} and $3\sigma$ 642 nm upper limits for LkCa 15 b, respectively. The lines show accretion disc and hot-start models. The disc models are simple combinations of blackbody spectra\cite{2015ApJ...803L...4E}, a suitable approximation for the case of a cool (T $<$ 1500 K) stellar atmosphere where dust opacity dominates. The $M_p \dot{M}$ calculated from the H$\alpha$ flux agrees with that inferred from the infrared measurements (see text).}
\label{fig-sed}
\end{figure}

\renewcommand{\figurename}{\textbf{Extended Data Table}}
\setcounter{figure}{0}

\begin{figure*}
\includegraphics[width=\textwidth]{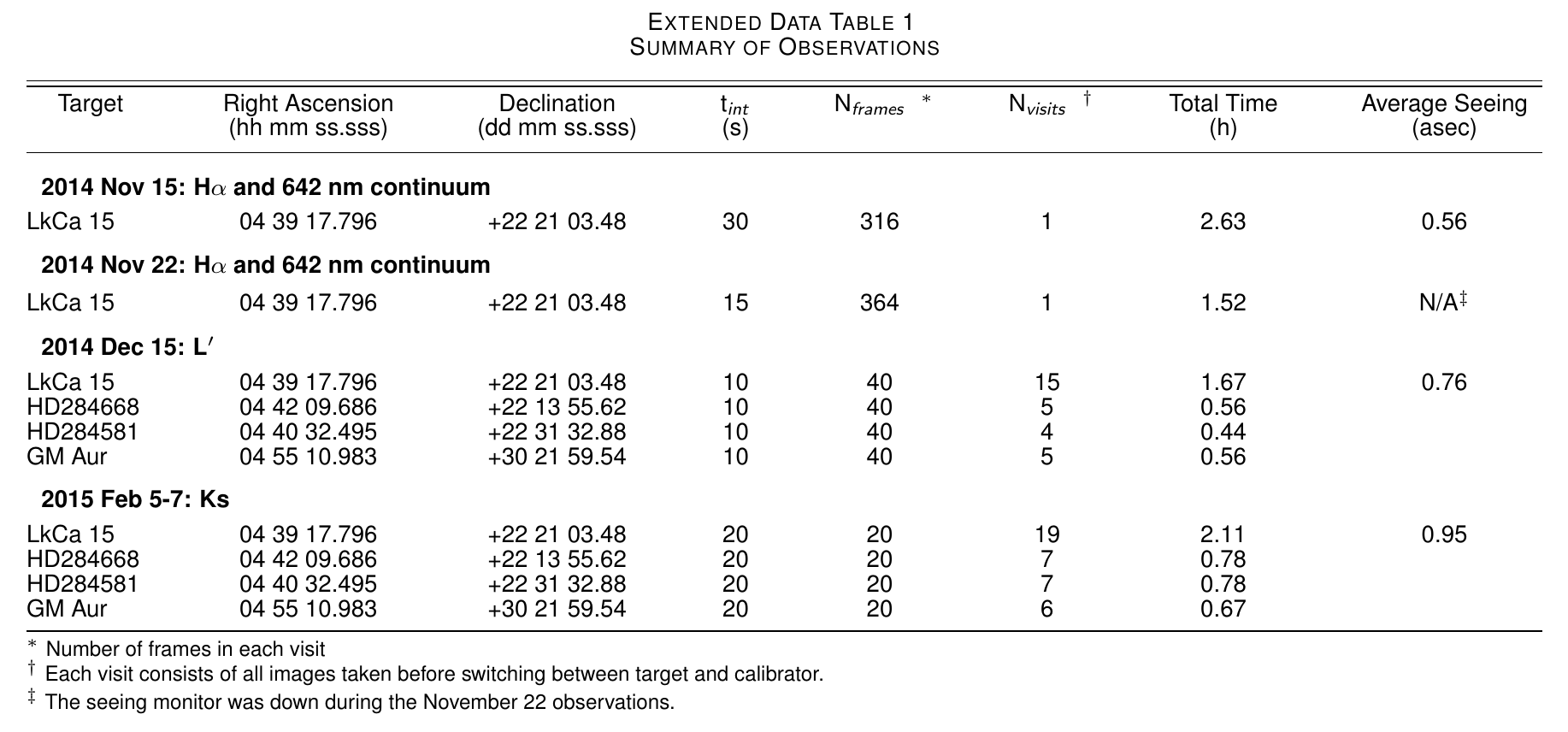}
\end{figure*}

\begin{figure*}
\includegraphics[width=0.6\textwidth]{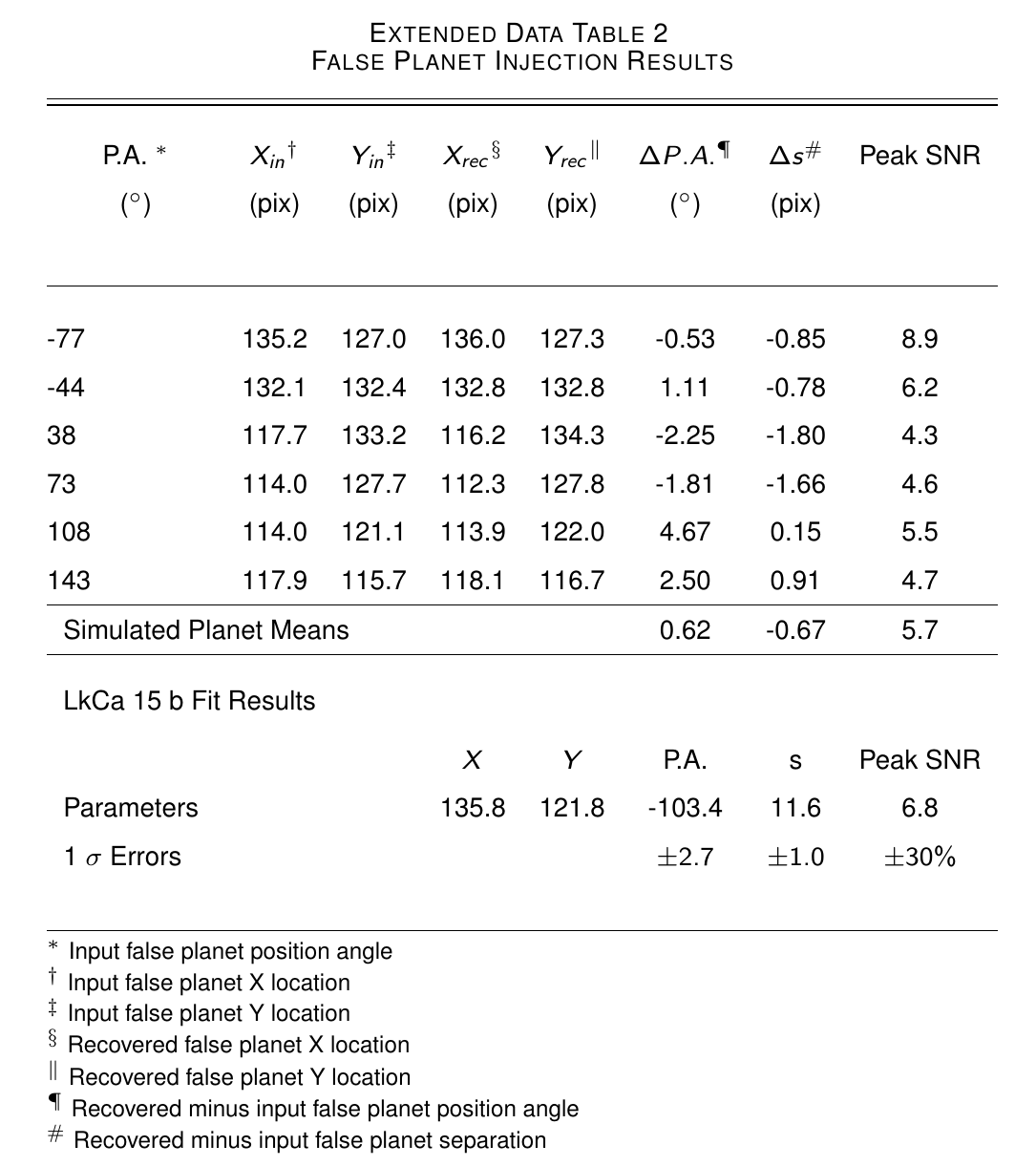}
\end{figure*}

\renewcommand{\figurename}{\textbf{Extended Data Figure}}
\setcounter{figure}{0}
\begin{figure*}
\includegraphics[width=\textwidth]{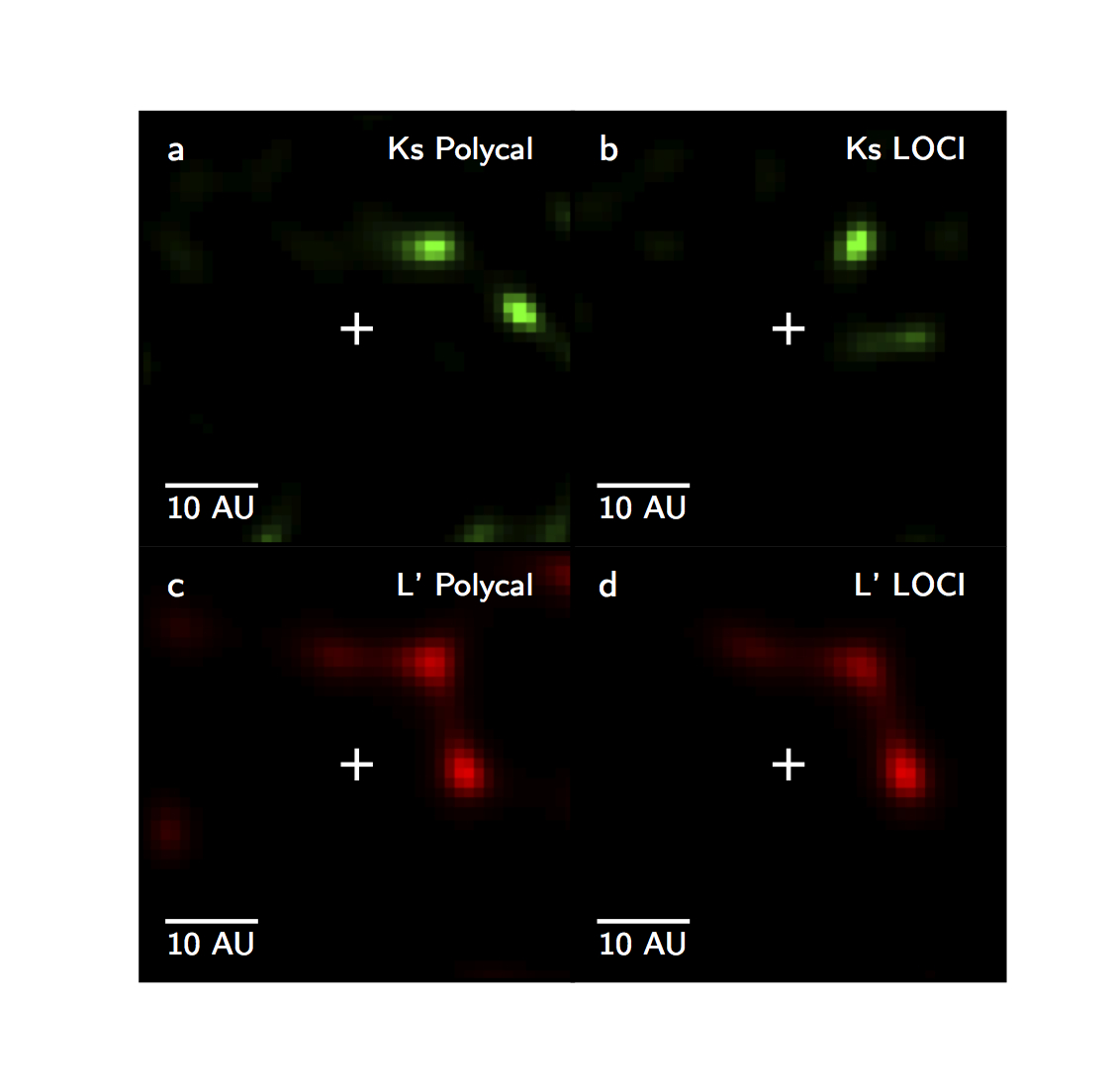}
\caption{\textbf{Image reconstructions.} Images reconstructed from closure phases, showing Ks polynomial (\textbf{a}) and LOCI-like (\textbf{b}) calibrations, and L$^\prime$ polynomial (\textbf{c}) and LOCI-like (\textbf{d}) calibrations. Both calibrations yielded reconstructed images with at least two distinct components. The LOCI-like calibration moved each companion within the position errors derived from the grid $\chi^2$ surface.}
\end{figure*}

\begin{figure*}
\includegraphics[width=\textwidth]{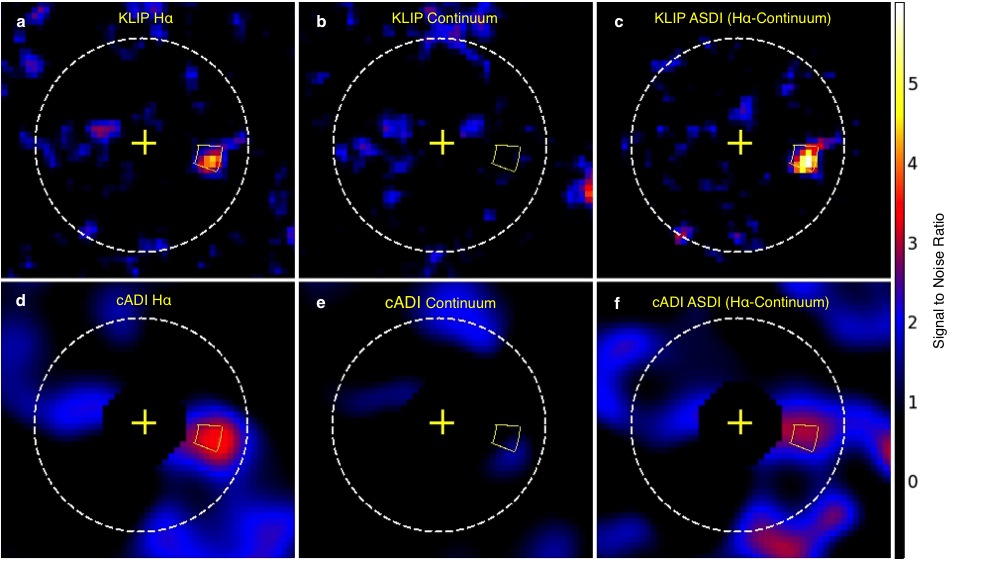}
\caption{\textbf{KLIP and ADI H$\alpha$ SNR maps.} \textbf{a-c)} Final KLIP SNR maps for H$\alpha$ (\textbf{a}), continuum (\textbf{b}) and the difference between the two (ASDI, \textbf{c}). \textbf{d-f)} Final cADI SNR maps in the same order. Dividing by the radial noise profiles to create these maps should normalize the noise distribution at all radii within the speckle-dominated regime. The presence of dark holes in the maps suggests that we are speckle-dominated out to the AO control radius at $r\sim20$ pixels (white, dashed circles). LkCa 15 b's separation is 11.6 pixels. The yellow keystones indicate the 2-sigma range of allowed astrometry for the KLIP ASDI point source (upper right) based on negative simulated planet injection.}
\end{figure*}

\begin{figure*}
\includegraphics[width=\textwidth]{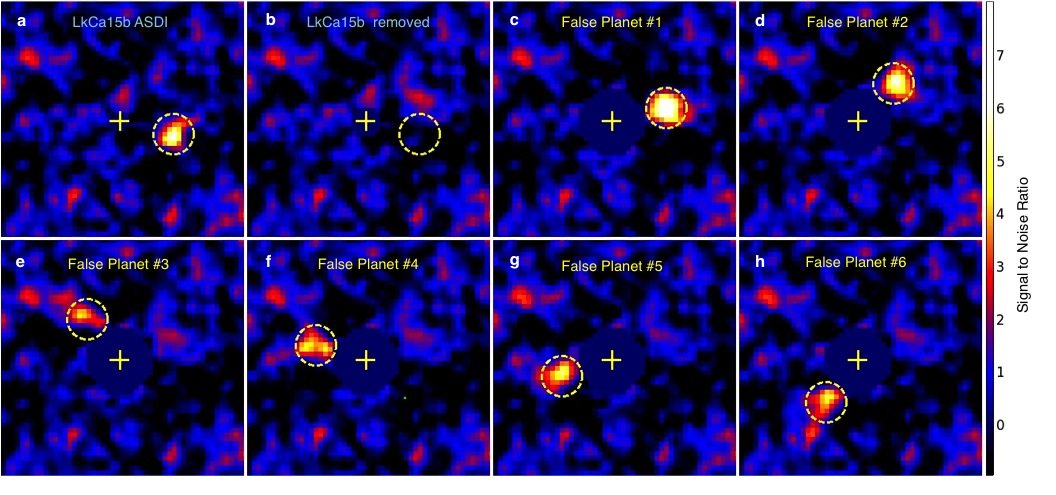}
\caption{\textbf{False positive planet SNR maps.} \textbf{a)} LkCa 15 final ASDI SNR map. \textbf{b)} ASDI SNR map with LkCa 15 b removed. \textbf{c-h)} ASDI SNR maps of false positive planets injected at a radius of 11 pixels and contrast of $8\times10^{-3}$. Recovered parameters for these planets are given in Extended Data Table 2 and were used to determine 1$\sigma$ astrometric and photometric uncertainties.}
\end{figure*}

\begin{figure*}
\includegraphics[width=0.7\textwidth]{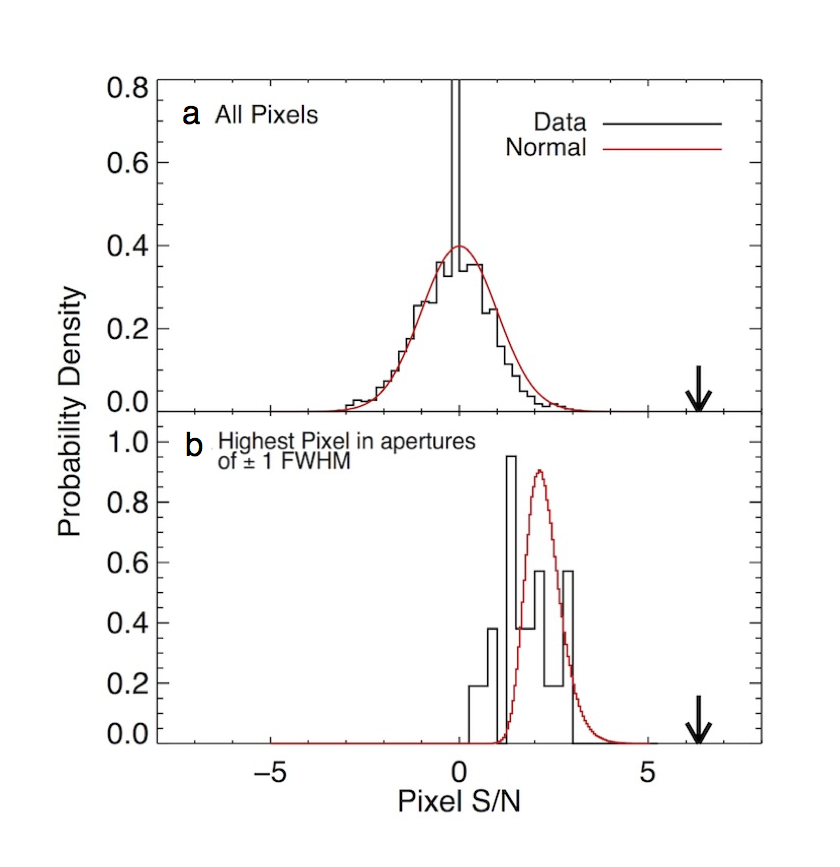}
\caption{\textbf{H$\alpha$ detection noise statistics.} \textbf{a)} Histogram of noise (non-planet) pixel values in the SNR map within the speckle dominated regime (black line) compared to a Normal distribution (red line). The black arrow denotes the location of the peak SNR value for LkCa 15 b. \textbf{b)} Histogram of the peak values in all noise apertures (see Extended Data Figure 5) within the control radius (black line) compared to a Normal distribution (red line). The black arrow shows the peak pixel value in the LkCa 15 b aperture.}
\end{figure*}

\begin{figure*}
\includegraphics[width=\textwidth]{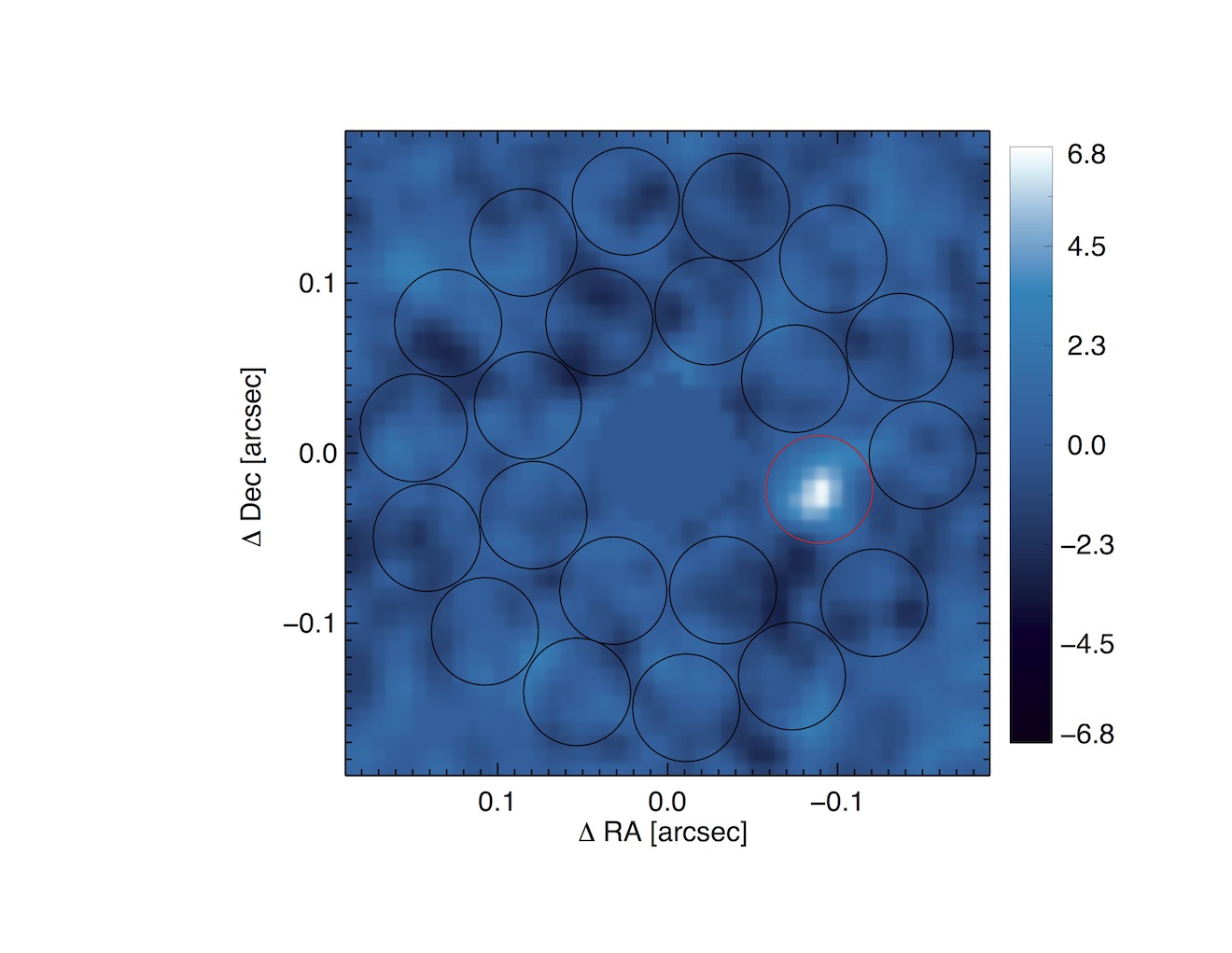}
\caption{\textbf{Noise apertures.} Noise apertures (black circles) surrounding LkCa 15 A used to calculate the statistics presented in Extended Data Figure 4. Color indicates SNR.}
\end{figure*}

\begin{figure*}
\includegraphics[width=0.7\textwidth]{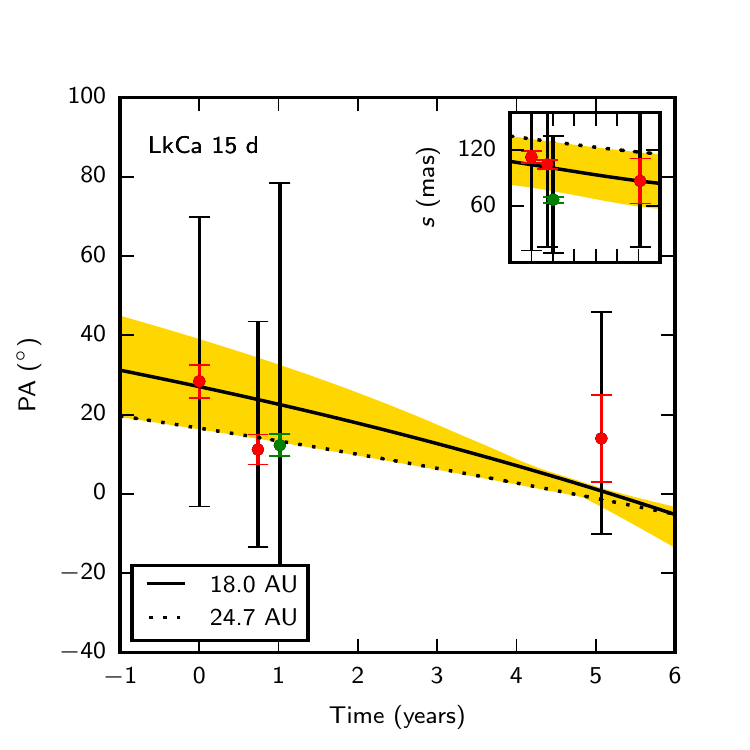}
\caption{\textbf{LkCa 15 d position angle and separation versus time.} Evolution of position angle and separation (inset) for LkCa 15 d. Green and red points indicate Ks and L$^\prime$ data, respectively. In both panels, the earliest three points correspond to previously published Keck observations,$^{8}$ and the most recent points show best fits to our data. The coloured error bars are derived using the non-linear algorithm MPFIT, which significantly underestimates the parameter errors compared to the more robust grid $\Delta \chi^2$ (black error bars; see Methods). The yellow shaded region spans the position angles and separations allowed at $1 \sigma$ by the multi-epoch observations, which have semi-major axes between 12.6 and 24.7 AU. Solid curves show the best-fit orbit (18.0 AU), and dashed curves show an orbit (24.7 AU) that is stable for a 0.5 $\mathrm{M_J}$ planet exterior to LkCa 15 b and c. Lower mass planets or resonant configurations permit stable orbits for LkCa 15 d at smaller stellocentric radii.
}
\label{fig-orbS}
\end{figure*}

\begin{figure*}
\includegraphics[width=\textwidth]{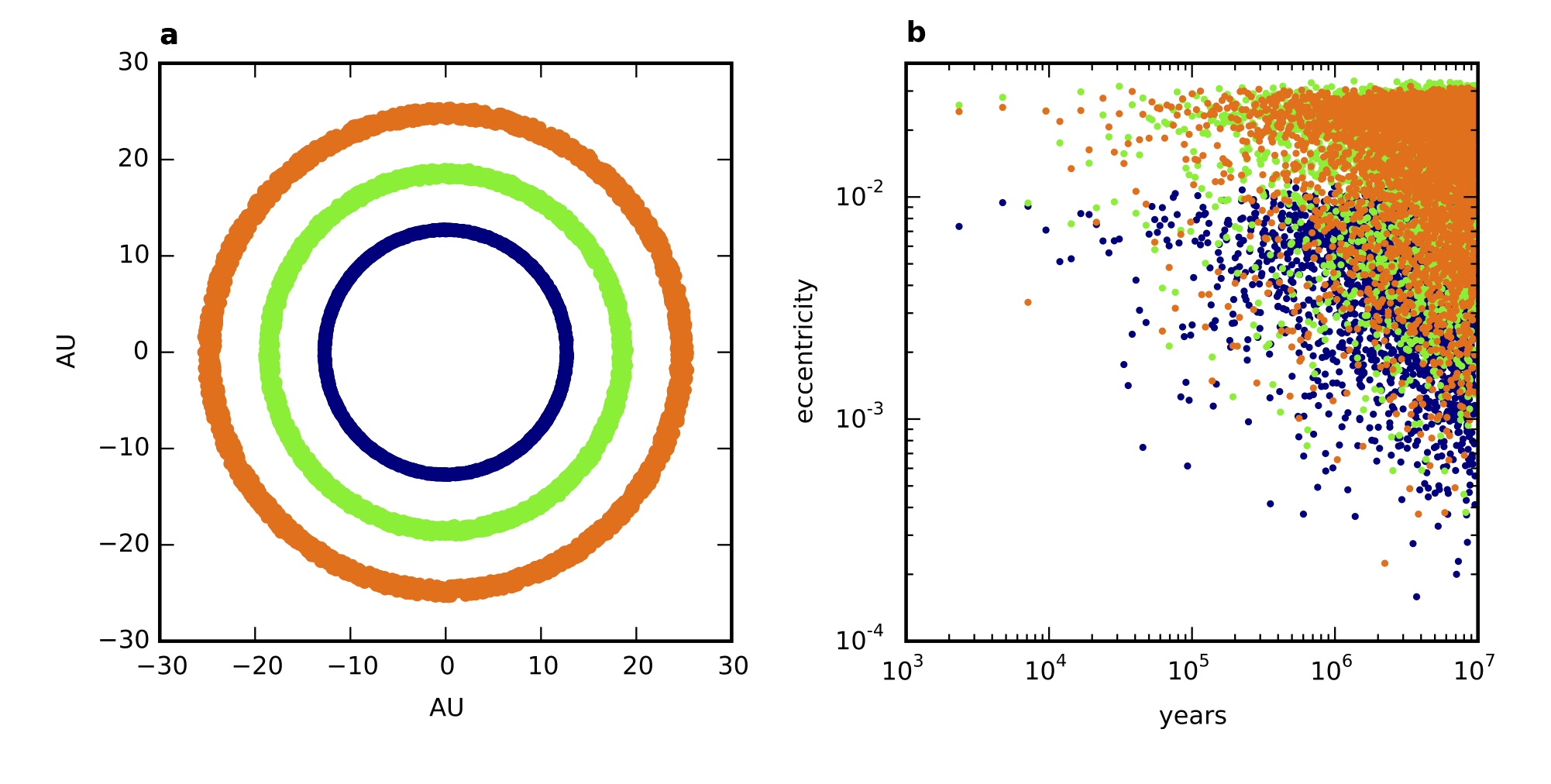}
\caption{\textbf{Orbital integration results.} \textbf{a)} Stable orbits for LkCa 15 b, c, and d over a 10 Myr integration. \textbf{b)} Osculating eccentricity. The planets are each 0.5 $\mathrm{M_J}$ with initial semi-major axes of 12.7, 18.6, and 24.7 AU, initial eccentricities of order $10^{-5}$, and relative inclinations of $< 1^\circ$. After a 10 Myr integration, the eccentricities of c and d have increased to only a few percent.}
\end{figure*}

\pagebreak



\end{document}